\documentclass[12pt]{article}
\usepackage{amssymb}
\usepackage{amsmath}
\usepackage{epsfig}
\usepackage{citesort}
\makeatletter
\renewcommand{\thefootnote}{\fnsymbol{footnote}}
\makeatother
\begin{document}
\thispagestyle{empty}
\begin{center}
{\Large
{\bf
Threshold resummation $S$-factor in QCD: \\
the case of unequal masses}}
\\
\vspace{5mm}
\normalsize
{\bf
O.P.~Solovtsova$^{\,a,b}$
and
Yu.D.~Chernichenko$^{\,a}$
}
\vskip 0.4em
{$^{a}${\it International Center for Advanced Studies, \\
Gomel State Technical University,
Gomel 246746, Belarus}}\\[0.2cm]
{$^{b}${\it Bogoliubov Laboratory of Theoretical Physics, \\
Joint Institute for Nuclear Research,
Dubna 141980, Russia}}

\vspace{5mm}

{\bf Abstract}\\[5mm]
{\parbox{16cm} {The new relativistic Coulomb-like threshold
resummation $S$-factor in quantum chromodynamics is obtained.
The consideration is given within the framework of quasipotential
approach in quantum field theory which is formulated in relativistic
configuration representation for two particles of unequal masses. }}
\end{center}

\setcounter{footnote}{0}
\renewcommand{\thefootnote}{\arabic{footnote}}

\renewcommand{\baselinestretch}{1.4}

\section{Introduction}
As is well known, a description of quark-antiquark systems close to
threshold does not permit us to cut off the perturbative series even
if the expansion parameter, the QCD coupling constant $\alpha_s$, is
small~\cite{App-Politzer75,PQW}. The reason consist in that the
real expansion parameter in the threshold region is $ \alpha/v\,$,
where $v=\sqrt{1-4{m^2}/s}$ is a quark velocity and $m$ is a quark
mass. The problem is well known from QED~\cite{Schwinger:73}.
To obtain a meaningful result, these threshold singularities of the
form $(\alpha/v)^n$ have to be summarized. In the nonrelativistic
case, for the Coulomb interaction
\begin{equation}\label{eq1}
V(r)=-\frac{\alpha}{r}
\end{equation}
this resummation is realized by the known
Gamov--Sommerfeld--Sakharov
$S$-factor~\cite{Gamov:28,Sommerfeld,Sakharov}
\begin{equation}
\label{eq2}
S_{\rm nr}=\frac{X_{\rm{nr}}}{1-\exp(-X_{\rm{nr}})}\,,
\quad\quad X_{\rm nr}=\frac{\pi\,\alpha}{v_{\rm{nr}}}\,,
\end{equation}
which is related to the wave function of the continuous spectrum at
the origin by $|\psi(0)|^2$ (see books \cite{Goldberger,Newton}).
Here $2\,v_{\rm{nr}}$ is the relative velocity of two nonrelativistic
particles. The corresponding nonrelativistic expression can also be
obtained for higher $\ell$ states (see, e.g.,
Ref.~\cite{Adel-Yndurain:95}).

In the relativistic theory the nonrelativistic approximation needs to
be modified. The relativistic modification of the
$S$-factor~(\ref{eq2}) in QCD in the case of two particles of equal
masses $(m_1=m_2=m)$ was executed in Ref.~\cite{Fadin88} (see also
Ref.~\cite{Fadin95}) and it consisted in the change
$v_{\rm{nr}}\rightarrow\,v$. This factor was used for the description
of effects close to the threshold of pair production in the processes
$e^{+}e^{-}\to\,t\bar{t}\,$ and
$\,e^{+}e^{-}\to\,W^{+}W^{-}$~\cite{Fadin95}. Just the same form of
the $S$-factor for the interaction of two particles of equal masses
was later suggested in Ref.~\cite{Hoang97}. Another form of the
relativistic generalization of the $S$-factor also in the case of two
particles of equal masses was obtained in Ref.~\cite{YoonWong:0005}.
The relativistic $S$-factor for two particles of arbitrary masses
($m_1\ne\,m_2$) we are interested in was presented in
Ref.~\cite{Arbuzov94}. This factor was derived within the framework
of relativistic quantum mechanics on the basis of the Schr\"odinger
equation.

The new step to relativistic generalization of the $S$-factor in the
case of two particles of equal masses was made by Milton and
Solovtsov in Ref.~\cite{Milton-Solovtsov_ModPL:01}. For this purpose,
the relativistic quasipotential (RQP) approach proposed by Logunov
and Tavkhelidze~\cite{LogunovTav} in the form suggested by
Kadyshevsky~\cite{Kadyshevsky68} turned out to be convenient. In
Ref.~\cite{Milton-Solovtsov_ModPL:01}, a transformation of the
quasipotential (QP) equation from momentum space into a relativistic
configuration representation for two particles of equal masses (see
Ref.~\cite{KadMS:68}) was used. Moreover, in
Ref.~\cite{Milton-Solovtsov_ModPL:01}, the potential~(\ref{eq1})
considered in Ref.~\cite{SavrinS80} was used which takes into account
its QCD-like behaviour. The solution containing arbitrary functions
of $r$ with period $i$, the so-called $i$-periodic constants, with
the same potential was investigated in Ref.~\cite{Freeman69}.
However, the use of this kind of solution is adequate for the
spectral problems only. Other forms of the QP equation with the
Coulomb potential were considered in Ref.~\cite{KapshaiSk8386}.

Thus, in Ref.~\cite{Milton-Solovtsov_ModPL:01}, a new step to
application of the quasipotential approach in QCD was made.
This approach gives the following expression for the
relativistic $S$-factor:
\begin{equation}
\label{eq3}
S(\chi)=\frac{X(\chi)}{1-\exp\left[-X(\chi)\right]}\,,\quad\quad
X(\chi)=\frac{\pi\,\alpha}{\sinh\chi}\,,
\end{equation}
where $\chi$ is the rapidity related to the total c.~m.~energy
of interacting particles $\sqrt{s}$ by
$2m\cosh\chi$ $=\sqrt{s}$.
The function $X(\chi)$ in Eq.~(\ref{eq3}) can be expressed in
terms of $v$ as $X(\chi)=\pi\,\alpha\sqrt{1-v^2}/v.$

The resummation factor appears in the parametrization of the
imaginary part of the quark current correlator, the Drell ratio
$R(s)$, which can be approximated in terms of the Bethe-Salpeter (BS)
amplitude of two charged particles $\chi_{\rm BS}(x)$ at $x=0$ (see
Ref.~\cite{BarbieriCR73}). The nonrelativistic replacement of this
amplitude by the wave function, which obeys the Schr\"odinger
equation with the Coulomb potential (\ref{eq1}), leads to
formula~(\ref{eq2}) with a substitution
$\alpha\rightarrow4\,\alpha_{s}/3$ for the QCD case. The possibility
of using the QP approach for our task is based on the fact that the
BS amplitude, which parameterizes the physical quantity $R(s)$, is
taken at $x=0$; therefore, in particular, at the relative time
$\tau=0$. Thus, the QP wave function in the momentum space is defined
as the BS amplitude at $\tau=0$ and, therefore, $R(s)$ can be
expressed in terms of the QP wave function in the momentum space,
$\Psi_{q}({\bf p})$, by using the relation
\begin{equation}
\label{eq4}
\chi_{\rm BS}(x=0)=\frac{1}{(2\pi)^3}\,\int
d\Omega_p\,\Psi_{q}(\mathbf{p})\, ,
\end{equation}
where $d\Omega_p=(m\,d{\mathbf{p}})/E_p$ is the relativistic
three-dimensional volume element in the Lobachevsky space realized on
the hyperboloid $E_p^2-{\mathbf{p}}^2=m^2$.

The relativistic $P$-factor (for $\ell=1$ state) in the case of two
particles of equal masses was obtained in Ref.~\cite{SSCh:05}. In that
paper, a new model expression for $R(s)$, in which threshold
singularities were summarized to the main potential contribution,
was suggested as well.

The generalization of the relativistic $S$- and $P$-factors for
arbitrary $\ell$ states in the case of two particles of equal masses
was discussed in Ref.~\cite{SCh:07}. Applications of the
factor~(\ref{eq3}) for describing some hadronic processes can be
found in Refs.~\cite{MSS_Adler:01,SS_NonLin:02,MSS:06}. Recently, the
relativistic $S$-factor~(\ref{eq3}) has been applied to reanalyze the
mass limits obtained for magnetic monopoles which might have been
produced at the Fermilab Tevatron~\cite{Milton:08}.

The purpose of this paper is to generalize the  previous study started
in Ref.~\cite{Milton-Solovtsov_ModPL:01} to the case of the interaction
of two particles of unequal masses ($m_{1}\neq m_{2}$).
The paper is organized as follows. In the next section, we present
the formalism of the RQP approach in quantum field theory formulated in
the relativistic configuration representation for the interaction of
two relativistic particles of unequal masses.
In Sec.~III, within the framework of this approach we derive the new
relativistic $S$-factor and analyze its behavior in the following
cases: the nonrelativistic and relativistic cases, the case of equal
masses, the case of one particle being at rest, and the
ultrarelativistic case.
Also, we compare the obtained factor with the factors considered in
Refs.~\cite{Fadin88,Fadin95,Hoang97,YoonWong:0005,Arbuzov94} and
study the behavior of the function $R(s)$ that is expressed in terms
of these factors in the case of vector current for two quarks of
unequal masses. Summarizing comments are given in Sec.~IV.

\section{The integral form of the quasipotential equation:\\[-0.2cm]
 the case of two particles of unequal masses}
A starting point of our consideration is the QP equation in the
momentum space constructed in Ref.~\cite{KadMM:70} for the RQP wave
function $\Psi_{q'}({\bf p}')$ of two relativistic particles of
unequal masses. This equation is given by \footnote{In the following
we will use the system of units $\hbar=c=1$.}
\begin{equation}
\label{eq5}
\left(2E_{q'}\,-\,2E_{p'}\right)\,\Psi_{q'}({\bf p}')
=\frac{2\,\mu}{m'\,(2\pi)^3}\,\int\,d\Omega_{k'}\,\widetilde{V}
\left({\bf p}'\,,{\bf k}';\,E_{q'}\right)\,\Psi_{q'}({\bf k}')\,,
\end{equation}
where
\[
d\Omega_{k'}=\frac{m'\,d{\bf k}'}{E_{k'}}
\]
is the relativistic three-dimensional volume element in the
Lobachevsky space, $E_{k'}=\sqrt{m'\,^2+{\bf k}'\,^2}$,\,
$m'=\sqrt{m_1\,m_2}$, and $\mu=m_1\,m_2/(m_1+m_2)$ is the usual
reduced mass. \footnote{Various definitions of the relativistic
reduced mass were discussed in Ref.~\cite{Faustov85}.}

Equation~(\ref{eq5}) represents a relativistic generalization of the
Lippmann-Schwinger equation in the spirit of the Lobachevsky geometry
which is realized on the upper half of the mass hyperboloid
$E_{k'}^2-{\bf k}'\,^2=m'\,^2$. This equation describes the
scattering over the quasipotential $\widetilde{V}\left({\bf
p}',\,{\bf k}';\,E_{q'}\right)$ of an effective relativistic particle
having mass $m'$ and a relative 3-momentum ${\bf k}'$, emerging
instead of the system of two particles and carrying the total c.~m.
energy of the interacting particles, $\sqrt{s}$, proportional to the
energy $E_{k'}$ of one effective relativistic particle of mass
$m'$~(see \cite{KadMM:70,KadMS:72}):
\begin{equation}
\label{eq6}
\sqrt{s}=\sqrt{{m_1}^2+{\bf k}^2}+\sqrt{{m_2}^2+
{\bf k}^2}=\frac{m'}{\mu}\,E_{k'},\,\quad E_{k'}=\sqrt{m'\,^2+
{\bf k}'\,^2}.
\end{equation}
The proper Lorentz transformations mean a translation in the
Lobachevsky space. The role of the plane waves corresponding to these
translations is played  by the following functions:
\begin{equation}
\label{eq7}
\xi(\mathbf{p}'\,,\mathbf{r})=
\left(\frac{E_{p'}-\mathbf{p}'\cdot\mathbf{n}}{m'}\right)^{-1-ir\,m'}\,,
\end{equation}
where the module of the radius-vector, $\bf{r}$,
($\mathbf{r}=r\,\bf{n}\,,|\mathbf{n}|=1$) is a relativistic
invariant~\cite{KadMS:72}. These functions correspond to the
principal series of unitary representations of the Lorentz group and
in the nonrelativistic limit ($p'\ll1$, $r\gg1$) $\xi(\mathbf{p}'\,,
\mathbf{r})\to\exp(i\mathbf{p}'\cdot\mathbf{r})$. The
functions~(\ref{eq7}) obey the following conditions of completeness
and orthogonality~\cite{KadMS:72}:
\begin{equation}
\label{eq8}
\begin{array}{c}
\displaystyle\frac{1}{(2\pi)^3}\,
\int d\Omega_{p'}\,\xi(\mathbf{p}'\,,
\mathbf{r})\,\xi^*(\mathbf{p}'\,,\mathbf{r}')=
\delta(\mathbf{r}-\mathbf{r}')\,,\\[0.5cm]
\displaystyle\frac{1}{(2\pi)^3}\,\int
d\mathbf{r}\,\xi(\mathbf{p}'\,,\mathbf{r})\,
\xi^*(\mathbf{q}'\,,\mathbf{r})=
\delta(\mathbf{p}'(-)\mathbf{q}')\,,
\end{array}
\end{equation}
where $\delta(\mathbf{p}'(-)\mathbf{q}')=
\sqrt{1+\mathbf{p}'\,^2/{m'\,^2}}\,\delta(\mathbf{p}'-\mathbf{q}')$
is the relativistic $\delta$-function in momentum-space. Moreover,
these the functions satisfy the equation in terms of finite
differences
\begin{equation}
\label{eq9}
\left(2E_{p'}\,-\,\hat{H}_0\right)\,\xi({\bf p}'\,,{\bf r})=0\,.
\end{equation}
Here
\begin{equation}
\label{eq10}
\hat{H}_{0}=2\,m'\,\left[\cosh\left(i\lambda'
\frac{\partial}{\partial{r}}\right)+\frac{i\lambda'}{r}
\sinh\left(i\lambda'\frac{\partial}{\partial{r}}\right)-
\frac{\lambda'\,^2\,\Delta_{\theta,\varphi}}{2r^2}
\exp\left(i\lambda'\frac{\partial}{\partial{r}}\right)\right]
\end{equation}
is the operator of the free Hamiltonian, while
$\Delta_{\theta,\varphi}$ is its angular part and $\lambda'=1/m'$ is
the Compton wavelength associated with the effective relativistic
particle of mass $m'$.

The QP wave functions in the momentum space and relativistic
configuration representation~\cite{KadMM:70,KadMS:72} are related
as follows:
\begin{equation}
\label{eq11}
\begin{array}{c}
\psi_{q'}(\mathbf{r})=\displaystyle\frac{1}{(2\pi)^3}\,
\int d\Omega_{p'}\,\xi(\mathbf{p}'\,,\mathbf{r})\,
\Psi_{q'}(\mathbf{p'})\,,\\[0.7cm]
~\Psi_{q'}({\mathbf{p}'})=\displaystyle\int d\mathbf{r}\,
\xi^*(\mathbf{p}'\,,\mathbf{r})\,\psi_{q'}(\mathbf{r}).~~~~~~~~~~~~
\end{array}
\end{equation}
For a spherically symmetric potential the application of
transformations~(\ref{eq11}) (Shapiro transformations or
$\xi$-transformations) to Eq.~(\ref{eq5}) leads to the equation which
is the integral form of the relativistic Schr\"odinger equation in
the configuration representation:
\begin{equation}
\label{eq12}
\begin{array}{c}
\displaystyle\frac{1}{(2\pi)^3}\int\,d{\Omega_{p'}}\,
\left(2E_{q'} -2E_{p'}\right)\,\xi(\mathbf{p}'\,,\mathbf{r})
\int\,d{\mathbf{r}'}\,\xi^*(\mathbf{p}'\,,\mathbf{r}')\,
\psi_{q'}({\bf r}')=\displaystyle\frac{2\mu}{m'}\,V(r)\,
\psi_{q'}({\bf r})\,,
\end{array}
\end{equation}
where the right-hand side is already local in the configuration
representation and the transform of the potential, $V(r)$, is given
in terms of the same relativistic plane waves.

We note that the use relations~(\ref{eq11}) and
Eq.~(\ref{eq9}) allows us to express the left-hand side of
Eq.~(\ref{eq12}) in terms finite differences
\begin{equation}
\label{eq13}
\left(2E_{q'}-\hat{H}_0\right)\psi_{q'}({\bf r})=
\frac{2\,\mu}{m'}\,V(r)\,\psi_{q'}({\bf r})\,.
\end{equation}
Solutions of this equation, in principle, can contain arbitrary
functions of $r$ with period $i$, the so-called $i$-periodic
constants, which appear in the solutions due to the finite difference
nature of the Hamiltonian~(\ref{eq10}). For some problems, such as
defining the bound state spectrum, this $i$-periodic constant is not
important. However, for the purpose of extracting resummation factors
one must develop a method which avoids this ambiguity. For this
purpose instead of Eq.~(\ref{eq13}) we will to use Eq.~(\ref{eq12}).
This equation can be reduced to the form
\begin{equation}
\label{eq14}
\displaystyle\frac{1}{(2\pi)^3}\,
\int d{\Omega_{p}}\,\left(2E_{q} -2E_{p}\right)\,\xi(\mathbf{p}\,,
\boldsymbol{\rho})\,
\int d{\boldsymbol{\rho}'}\,\xi^*(\mathbf{p}\,,
\boldsymbol{\rho}')\,\psi_{q}(\boldsymbol{\rho}')=
\frac{2\,\mu}{m'}\,V(\rho)\,\psi_{q}(\boldsymbol{\rho})\,.
\end{equation}
We introduced the following notations:
\begin{equation}
\label{eq15}
\begin{array}{c}
\mathbf{q}'=m'\,\mathbf{q}\,,\mathbf{p}'=m'\,\mathbf{p}\,,
\mathbf{q}=\sinh(\chi_q)\,\mathbf{n}_{q}\,,
\mathbf{p}=\sinh(\chi_{p})\,\mathbf{n}_p\,,|\mathbf{n}_
{q}|=|\mathbf{n}_{p}|=1\,,\\[0.4cm]
\boldsymbol{\rho}=m'\,\mathbf{r}\,,\boldsymbol{\rho}'=m'\,
\mathbf{r}'\,,\,\rho=m'\,r\,,\,\rho'=m'\,r'\,,\,\\[0.2cm]
\displaystyle d\mathbf{r}'={m'}^{-3}\,d\boldsymbol{\rho}'\,,
d\Omega_{p'}={m'}^3\,d\Omega_{p}\,,d\Omega_{p}=
\frac{d\mathbf{p}}{E_{p}}\,,E_{q'}=m'\,E_{q}\,,E_{p'}=m'
\,E_{p}\,,\\[0.5cm]
E_{q}=\sqrt{1+{{\bf q}}^2}\,,E_{p}=\sqrt{1+{{\bf p}}^2}\,,\,
\xi(\mathbf{p}'\,,\mathbf{r})=\displaystyle\left(E_{p}-
{\bf p}\cdot{\bf n}\right)^{-1-i\rho}\equiv\xi(\mathbf{p}\,,
\boldsymbol{\rho})\,,\\[0.5cm]
V(r)=V(\rho/m')\equiv m'\,V(\rho)\,,\psi_{q'}(\mathbf{r})=
\psi_{m'\,q}(\boldsymbol{\rho}/m')\equiv\psi_{q}
(\boldsymbol{\rho})\,,\Psi_{q'}(\mathbf{p}')\equiv{m'}^{-3}\,
\Psi_{q}(\bf{p})\,.
\end{array}
\end{equation}

By using the expansions
\begin{equation}
\label{eqdop1}
\begin{array}{c}
\xi(\mathbf{p}\,,\boldsymbol{\rho})=\displaystyle\sum_{\ell=0}^{\infty}\,(2\ell+1)\,i^{\ell}\,p_{\ell}(\rho\,,
\cosh\chi_p)\,P_{\ell}\biggl(\frac{\mathbf{p}\cdot\boldsymbol{\rho}}{p\,\rho}\biggr)\,,\\[0.6cm]
\psi_{q}(\boldsymbol{\rho})=\displaystyle\sum_{\ell=0}^{\infty}\,(2\ell+1)\,i^{\ell}\,\frac{\varphi_{\ell}(\rho\,,
\chi_q)}{\rho}\,P_{\ell}\biggl(\frac{\mathbf{q}\cdot\boldsymbol{\rho}}{q\,\rho}\biggr)\,,
\end{array}
\end{equation}
and also formula (see Ref.~\cite{KadMS:68})
\[
p_{\ell}(\rho\,,\cosh\chi)=\frac{(-1)^{\ell}\,(\sinh\chi)^{\ell}}{\rho^{(\ell+1)}}\biggl(\frac{d}
{d\cosh\chi}\biggr)^{\ell}\biggl(\frac{\sin \rho\chi}{\sinh\chi}\biggr)\,,
\]
Eq.~(\ref{eq14}) is transformed to
\begin{equation}
\label{eq16}
\begin{array}{c}
\displaystyle\frac{2}{\pi}\int\limits_{0}^{\infty}d\chi'\frac{(\sinh\chi')^{2\ell+2}\,(-1)^{\ell+1}}{\rho^{(\ell+1)}}
\left(2\cosh\chi-2\cosh\chi'\right)\biggl(\frac{d}{d\cosh\chi'}\biggr)^{\ell}\biggl(\frac{\sin \rho\chi'}
{\sinh\chi'}\biggr)\times\\[0.6cm]
\times\displaystyle\biggl(\frac{d}{d\cosh\chi'}\biggr)^{\ell}\frac{1}{\sinh\chi'}\int\limits_{0}^{\infty}d\rho'
\frac{\rho'\,\sin \rho'\chi'}{(-\rho')^{(\ell+1)}}\,\varphi_{\ell}(\rho'\,,\chi)=\frac{2\,\mu}{m'}\,
\frac{V(\rho)\,\varphi_{\ell}(\rho\,,\chi)}{\rho}\,.
\end{array}
\end{equation}
Here $P_{\mu}^{\nu}(z)$ is a Legendre function of the first kind, and
the function
\begin{displaymath}
p_{\ell}(\rho\,,\cosh\chi)=\frac{(-1)^{\ell+1}}
{\rho}\sqrt\frac{\pi}{2\sinh\chi}\,(-\rho)^{(\ell+1)}\,
P_{-1/2+i\rho}^{-1/2-\ell}(\cosh\chi)
\end{displaymath}
is the solution of Eq.~(\ref{eq13}) in the case when the
interaction is switched off, $V(r)\equiv0$; $\chi'$ and $\chi$ are the
rapidities which are related to $E_p\,,E_q$ as
$E_p=\cosh\chi'\,,E_q=\cosh\chi$, and the function
\begin{equation}
\label{eq17}
(-\rho)^{(\ell+1)}=i^{\ell+1}\frac{\Gamma(\ell+1+i\rho)}{\Gamma(i\rho)}
\end{equation}
is the generalized power~\cite{KadMS:68}
where $\Gamma(z)$ is the gamma-function.

Thus, Eq.~(\ref{eq16}) differs from the corresponding equation in the
case of two particles of equal masses (see \cite{SCh08}) only by the
factor $2\mu/m^{'}$ turning into 1 at $m_1=m_2$.

\section{Relativistic  $S$-factor}

The $\xi$-transformation~(\ref{eq11}) as applied to the Coulomb
interaction~(\ref{eq1}) gives the potential in momentum space
\[
V(\Delta)\,\sim\,\frac{1}{\chi_\Delta\sinh\chi_\Delta}\,,
\]
where the relative rapidity $\chi_\Delta$ corresponds to
${\mathbf\Delta}={\mathbf{p}'}(-){\mathbf{k}'}$ and is defined in
terms of the square of the momentum transfer by
$Q^2=-(p'-k')^2=2(\cosh\chi_\Delta-1)$ (see, for detail,
Ref.~\cite{Milton-Solovtsov_ModPL:01}). For large $Q^2$ the potential
$V(\Delta)$ behaves as $(Q^2\ln Q^2)^{-1}$, which reproduces the
principal behavior of the QCD potential proportional to
$\bar\alpha_{S}(Q^2)/Q^2$ with $\bar\alpha_{S}(Q^2)$ being the QCD
running coupling. This property of the potential~(\ref{eq1})
was noted in Ref.~\cite{SavrinS80}.

For the first time, the solution of Eq.~(\ref{eq16}) in the case
of the interaction of two relativistic particles of equal masses
at $\ell=0$, not containing the $i$-periodic constants, was
obtained in Ref.~\cite{Milton-Solovtsov_ModPL:01}. This approach leads
to the relativistic $S$-factor~(\ref{eq3}). To solve the RQP
equation~(\ref{eq16}) with the potential~(\ref{eq1}), we use the
method developed in Ref.~\cite{Milton-Solovtsov_ModPL:01} (see also
Refs.~\cite{SkSol83,SCh08}).

We will seek a solution of Eq.~(\ref{eq16}) with the
potential~(\ref{eq1}) in the form
\begin{equation}
\label{eq18}
\varphi_{\ell}(\rho\,,\chi)=\frac{(-\rho)^{
(\ell+1)}}{\rho}\int\limits_{\alpha}^{\beta}d\zeta\,e^{i\rho\zeta}\,
R_{\ell}(\zeta\,,\chi)\,,
\end{equation}
where the $\zeta$-integration is performed in the complex plane over
a contour with end points $\alpha$ and $\beta$:
$\alpha=-R-i\varepsilon\,,\beta=-R+i\varepsilon\,,R\to
+\infty\,,\varepsilon\to+0$.
Substituting~(\ref{eq18}) into~(\ref{eq16}) and taking into account
that
\[
\frac{1}{i\pi}\int\limits_{0}^{\infty}d\rho'
\sin(\rho'\chi')\,e^{i\rho'\zeta}=\frac{1}{i\pi}\frac{\chi'}{{\chi'}^2-
{\zeta}^2}\,,
\]
we obtain the equation
\begin{equation}
\label{eq19}
\begin{array}{c}
\displaystyle(-1)^{\ell}\int\limits_{\alpha}^{\beta}d\zeta\,R_{\ell}(\zeta\,,\chi)\biggl(\frac{d}{d\cosh\zeta}\biggr)^
{\ell}\biggl[(\sinh\zeta)^{2\ell+1}\,\left(2\cosh\chi-2\cosh\zeta\right)\times \\[0.7cm]
\displaystyle\times\biggl(\frac{d}{d\cosh\zeta}\biggr)^{\ell}\biggl(\frac{e^{i\rho\zeta}}{\sinh\zeta}\biggr)\biggr]=
-\frac{2\,\alpha\,\mu}{m'\,\rho}\prod\limits_{n=1}^{\ell}(\rho^2+n^2)\int\limits_{\alpha}^{\beta}d\zeta\,
e^{i\rho\zeta}\,R_{\ell}(\zeta\,,\chi)\,.
\end{array}
\end{equation}

It should be noted that solutions of this equation, and hence
Eq.~(\ref{eq16}), do not contain any more the $i$-periodic
constants which appear in the solutions of Eq.~(\ref{eq13}) due
to the finite difference nature of the Hamiltonian~(\ref{eq10}).

The solution of Eq.~(\ref{eq19}) at $\ell=0$ leads to the
following expression for the RQP partial wave function:
\begin{equation}
\label{eq20}
\displaystyle\varphi_0(\rho\,,\chi)=C_0(\chi)\frac{\rho}{\rho^{(1)}}\int\limits_{\alpha}^{\beta}d\zeta
\frac{e^{(i\rho+1)\zeta}}{(e^{\zeta}-e^{\chi})^2}\,\left[\frac{e^{\zeta}-e^{-\chi}}{e^{\zeta}-e^{\chi}}\right]^
{-1+iA}\,,
\end{equation}
where
\[
 A=\frac{\alpha\,\mu}{m'\,\sinh\chi}\,.
\]
Performing in Eq.~(\ref{eq20}) $\zeta$-integration in the complex plane
along a contour with end points $\alpha$ and $\beta$ (in the same way
as in Ref.~\cite{Milton-Solovtsov_ModPL:01,SkSol83,SCh08}) we obtain the
resulting solution which does not contain the $i$-periodic constant
in the form
\begin{equation}
\label{eq21}
\displaystyle\varphi_0(\rho\,,\chi)= C_0(\chi)\frac{2\,\rho\,\sinh(\pi\,\rho)}{\rho^{(1)}}\int\limits_{-\infty}^
{\infty}dx\frac{e^{(i\rho+1)x}}{(e^{x}+e^{\chi})^2}\left[\frac{e^{x}+e^{-\chi}}{e^{x}+e^{\chi}}\right]^{-1+iA}\,.
\end{equation}
This solution can also be represented in terms of
hypergeometric function as
\begin{equation}
\label{eq22}
\displaystyle\varphi_0(\rho\,,\chi)=-N_0(\chi)(-\rho)^{(1)}\,e^{i\rho\chi+iA\chi}\,
F\left(1-iA,1-i\rho;2;1-e^{-2\chi}\right)\,.
\end{equation}
The normalization constant $N_0(\chi)$ in Eq.~(\ref{eq22}) can be
obtained (see Ref.~\cite{Milton-Solovtsov_ModPL:01}) from the
condition
\begin{equation}
\label{eq23}
\lim_{\alpha \to 0}\varphi_{0}(\rho\,,\chi)=
\rho\,p_{0}(\rho\,,\cosh\chi)\xrightarrow[\rho\to\infty]\,
\frac{\sin(\rho\chi)}{\sinh\chi}\,.
\end{equation}

We should like to remind that the Bethe-Salpeter amplitude $\chi_{\rm
BS}(x=0)$ is associated with the RQP wave function in the momentum
space, $\Psi_q({\bf{p}})$, by relation~(\ref{eq4}). Taking into
account the transformations~(\ref{eq11}) and
notations (\ref{eq15}), the relationship of the Bethe-Salpeter
amplitude with the RQP wave function, $\psi_q(\boldsymbol{\rho})$, is
$$
\chi_{\rm{BS}}(x=0)=\psi_q(\boldsymbol{\rho})|_{\rho=i}\,.
$$
The generalized power~(\ref{eq17}) in the solution~(\ref{eq18})
vanishes at $\rho=i$ for all $\ell\ne0$. Thus, the
expansion~(\ref{eqdop1}) for the wave function
$\psi_q(\boldsymbol{\rho})$ contains only $s$-waves ($\ell=0$).
Hence, by using relations~(\ref{eq22}) and~(\ref{eq23}) we can
calculate $|\psi_q(i)|^2$, which leads to the following expression
for the relativistic $S$-factor in the case of two particles of
unequal masses:
\begin{equation}
\label{eq24}
S_{\rm\,uneq}(\chi)=\lim_{\rho\to{i}}
\left|\frac{\varphi_0(\rho\,,\chi)}{\rho}\right|^2=\frac{X_{\rm\,uneq}
(\chi)}{1-\exp\left[-X_{\rm\,uneq}(\chi)\right]}\,,
\quad\quad\,X_{\rm\,uneq}(\chi)=\frac{2\,\pi\,\alpha\,\mu}
{m'\,\sinh\chi}\,,
\end{equation}
where $\chi$ is the rapidity which is related to the total c.~m. energy,
$\sqrt{s}\,$, as $({m'}^2/\mu)\,\cosh\chi=\sqrt{s}$.

The function $X_{\rm\,uneq}(\chi)$ in Eq.~(\ref{eq24}) can be expressed
in terms of the ``velocity" $u$ determined by the relation
\begin{equation}
\label{eq25}
u=\sqrt{1-\frac{4\,{m'}^2}{\bar s}}\,,\quad\quad\,\bar s=s-(m_1-m_2)^2\,,
\end{equation}
in the form
\begin{equation}
\label{eq26}
X_{\rm\,uneq}(\chi)=\frac{\pi\,\alpha\,\sqrt{1-u^2}}{u}\,.
\end{equation}

We note that the square of relative 3-momentum ${\bf k}'$
for an effective relativistic particle, having mass $m'$, the total
c.~m. energy  of interacting particles, $\sqrt{s}$, and emerging
instead of the system of two particles, is defined by
formula~(\ref{eq6}) and is connected with the relativistic relative
velocity of interacting particles, $\rm\,v$, by the following
expression (see Refs.~\cite{KadMM:70,KadMS:72}):
\begin{equation}
\label{eq27}
{{\bf k}'}^2=\displaystyle2\mu^2\left({\frac{1}
{\sqrt{1-{\rm\,v}^2}}-1}\right)\,.
\end{equation}
In turn from relations~(\ref{eq6}) and~(\ref{eq27}) follows that the
relativistic relative velocity of interacting particles $\rm\,v$
can be expressed through their the total energy c.~m. of interacting
particles $\sqrt{s}$ by relation (exactly also as, for instance,
in Refs.~\cite{YoonWong:0005,Arbuzov94})
\begin{equation}
\label{eq28}
{\rm\,v}=\displaystyle2\,\sqrt
{\frac{s-(m_1+m_2)^2}{s-(m_1-m_2)^2}} \,\left[1+\frac{s-(m_1+m_2)^2}
{s-(m_1-m_2)^2}\right]^{-1}.
\end{equation}
Thence, taking into consideration the determination~(\ref{eq25}), we
find
\begin{equation}
\label{eq29}
{\rm\,v}=\frac{2\,u}{1+u^2}\,.
\end{equation}
Then expressions~(\ref{eq27}) and~(\ref{eq29}) give
\begin{equation}
\label{eq30}
{{\bf k}'}^2=(\mu)^2\,(u_{\rm rel}^{'})^2\,,
\end{equation}
where
\begin{equation}
\label{eq31}
u_{\rm rel}^{'}=\frac{2u}{\sqrt{1-u^2}}\,
\end{equation}
is the relative velocity of an effective relativistic particle with
mass $m'$ emerging instead of the system of two particles. This
result is found to be in full agrement with the physical meaning of
Eq.~(\ref{eq5}) which is a relativistic generalization of the
Lippmann-Schwinger equation in the spirit of Lobachevsky geometry.
This equation describes the scattering of an effective relativistic
particle on the quasipotential $\widetilde{V}\left({\bf p}',\,{\bf
k}';\,E_{q'}\right)$.
The effective particle emerges instead of the system of two
particles, has the mass $m'$, the relative 3-momentum ${\bf k}'$ and
carries the total c.~m. energy of interacting particles, $\sqrt{s}$.
Notice that the 3-momentum ${\bf k}'$ of an effective relativistic
particle and hence its relative velocity~(\ref{eq31}), according to
Eqs.~(\ref{eq27}) and~(\ref{eq30}), are invariants of the Lorentz
transformations.

Thus, in terms of relative velocity of an effective relativistic
particle (\ref{eq31}), the $S$-factor~(\ref{eq24}) is given by the
expression
\begin{equation}
\label{eq32}
\begin{array}{c}
\displaystyle S_{\rm\,uneq}(u_{\rm rel}^{'})=
\frac{X_{\rm\,uneq}(u_{\rm rel}^{'})}
{1-\exp\left[-X_{\rm\,uneq}(u_{\rm rel}^{'})\right]}\,,
\qquad X_{\rm\,uneq}(u_{\rm rel}^{'})=
\frac{2\,\pi\,\alpha}{u_{\rm rel}^{'}}\,.
\end{array}
\end{equation}
The factor~(\ref{eq32}) only formally has the same form as the
nonrelativistic factor~(\ref{eq2}). However, the factor~(\ref{eq32})
has an obviously relativistic nature since as the argument $r$ (the
module of radius-vector $\bf{r}$) in the Coulomb
potential~(\ref{eq1}) and the relativistic relative velocity of
interacting particles $\rm\,v$ (see Ref.~\cite{KadMS:72}) both are
relativistic invariants and hence the relative velocity of an
effective relativistic particle~(\ref{eq31}), according to
Eqs.~(\ref{eq27}) and~(\ref{eq30}), possesses this property as well.

The relativistic threshold resummation factor~(\ref{eq24})
[or of form~(\ref{eq32})] has the following important properties:

$\bullet$ In the nonrelativistic limit, $u\ll1$, it reproduces the
well-known nonrelativistic result.

$\bullet$ In the relativistic limit, $u\to1$, the
$S$-factor~(\ref{eq24}) [or~(\ref{eq32})] goes to unity.

$\bullet$ In the case of equal masses it coincides with the
$S$-factor~(\ref{eq3}).

$\bullet$ The case when one of the particles is at rest means that
$m_1\to\infty$. This gives the following limiting expression for
the ``velocity" $u$:
$$\displaystyle u\xrightarrow[m_1\to\infty]\,
\frac{|\mathbf{k}|}{\sqrt{{m_2}^2+{{\bf\,k}}^2}+m_2}\,.$$

$\bullet$ In the ultrarelativistic limit, as it was argued in
Refs.~\cite{Lucha:90,Lucha:96}, the bound state spectrum vanishes
since mass of an effective relativistic particle $m'\to0$. This
feature reflects an essential difference between potential models and
quantum field theory where an additional dimensional parameter
appears. One can conclude that within a potential model the
$S$-factor which corresponds to the continuous spectrum should go to
unity in the limit $m'\to0$. Thus, in contrast to the nonrelativistic
case, the relativistic resummation factor, the
$S$-factor~(\ref{eq24}) [or~(\ref{eq32})], reproduces both the known
nonrelativistic and the expected ultrarelativistic limits.

\begin{figure}[htb]
\centering\includegraphics[width=0.6\textwidth]{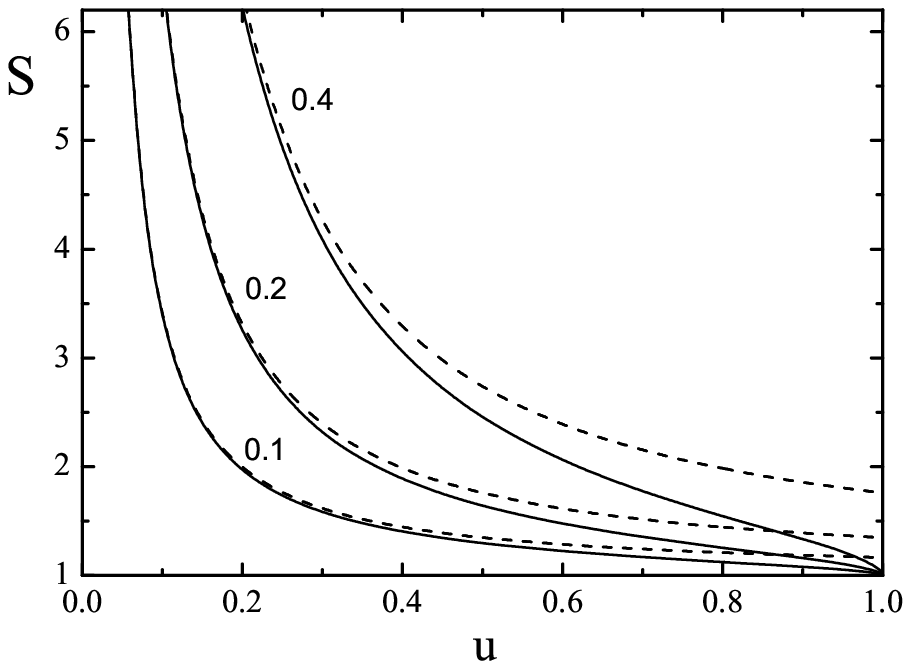}
\caption{Behavior of the $S$-factor at different values of the
constant $\alpha$ (the numbers at the curves). The solid lines
correspond to the new relativistic $S$-factor~(\ref{eq32}) and the
dashed lines to the nonrelativistic $S$-factor~(\ref{eq2}).}
\label{S_fig1}
\end{figure}

To illustrate the differences between the nonrelativistic
$S$-factor~(\ref{eq2}) and the new relativistic
$S$-factor~(\ref{eq32}) in more detail, in Fig.~\ref{S_fig1} we plot
the behavior of these factors as functions of $u$ at different values
of the constant $\alpha$ (the numbers at the curves). The solid lines
correspond to the $S$-factor~(\ref{eq32}) and the dashed lines to the
$S$-factor~(\ref{eq2}) with a substitution
$v_{\rm{nr}}\rightarrow\,u$. From this figure one can see that in the
region of nonrelativistic values of $u$, $u\leq0.2$, where the
influence of the $S$-factor is big, the difference between
(\ref{eq31}) and (\ref{eq2}) is practically absent. However, when
$\alpha$ increases, the nonrelativistic expression (\ref{eq2}) gives
a less suitable result in the region of large values $u$, in
particular, as $u\to1$.

It should be stressed that the $S$-factor~(\ref{eq32}) differs from
the $S$-factor for the interaction of two relativistic
particles of unequal masses
\begin{equation}
\label{eq33}
\begin{array}{c}
\displaystyle
S_{\rm\,A}({\rm\,v})=\frac{X_{\rm\,A}({\rm\,v})}
{1-\exp\left[-X_{\rm\,A}({\rm\,v})\right]}\,,
\qquad\displaystyle
X_{\rm\,A}({\rm\,v})=\frac{2\,\pi\,\alpha}{\rm\,v}\,,
\end{array}
\end{equation}
which was obtained in Ref.~\cite{Arbuzov94},
by the meaning of the relativistic relative velocity, $\rm\,v$, which
here is given by expression~(\ref{eq28}). Besides, in the case of
equal masses ($m_1=m_2=m$), the new $S$-factor~(\ref{eq32}) differs
also from the factor
\begin{equation}
\label{eq34}
\begin{array}{c}
\displaystyle S_{\rm\,H}({\rm\,v}_{\rm\,H})=\frac{X_{\rm\,H}({\rm\,v}_{\rm\,H})}{1-\exp\left[-X_{\rm\,H}
({\rm\,v}_{\rm\,H})\right]}\,,\,X_{\rm\,H}({\rm\,v}_{\rm\,H})=\frac{2\,\pi\,\alpha}{{\rm\,v}_{\rm\,H}}\,,\,
{\rm\,v}_{\rm\,H}=\frac{2\,\beta}{1+\beta^2}\,,\,\beta=\sqrt{1-\frac{4\,m^2}{s}}\,,
\end{array}
\end{equation}
which was presented in Ref.~\cite{Hoang97}.\footnote{The same
expression can be found in earlier paper~\cite{Fadin88}.} Indeed, the
factors~(\ref{eq33}) and~(\ref{eq34}) in form and in the
nonrelativistic limit (${\rm\,v}\,,{\rm\,v}_{\rm\,H}\,,u\to0$)
coincide with the factor~(\ref{eq32}). However, the relativistic
limits (${\rm\,v}\,, {\rm\,v}_{\rm\,H}\to1$) of the
factors~(\ref{eq33}) and~(\ref{eq34}) differ essentially from the
relativistic limit of the factor~(\ref{eq32}) equal to unity as
$u\to1$. Furthermore, in the case of the interaction of two
relativistic particles of equal masses the $S$-factor~(\ref{eq32})
differs from the factor
\begin{equation}
\label{eq35}
K=G(\eta)\,\kappa\,,
\end{equation}
where
\begin{equation}
\label{eq36}
G(\eta)=\frac{2\pi\eta}{1-\exp(-2\pi\eta)}\,,
\end{equation}
which was obtained in Ref.~\cite{YoonWong:0005}, not only in form but
also in a different behavior in the nonrelativistic
(${\rm\,v}\,,u\to0$) and the relativistic (${\rm\,v}\,,u\to1$) limits
(see Figs.~\ref{S_fig2} and \ref{S_fig3} below). The function
$G(\eta)$ in Eq.~(\ref{eq36}) is the nonrelativistic
$S$-factor~(\ref{eq2}) with $\eta=\alpha/{\rm\,v}$, and $\kappa$ is a
correction factor the expression for which contains the series and
infinite products as multipliers and is rather cumbersome. According
to Ref.~\cite{YoonWong:0005}, the influence of the $\kappa$ is
essential since $\kappa \to 1$ only as $\alpha\to 0$.

To illustrate the difference between the above factors, we show in
Fig.~\ref{S_fig2} the behavior of the  $S$-factors (\ref{eq2}),
(\ref{eq32}), (\ref{eq33}) and~(\ref{eq35}) as functions of $u$ for a
fixed value of ${\alpha} = 0.16$. The solid line represents the
relativistic $S$-factor~(\ref{eq32}); the dashed line the
nonrelativistic $S$-factor~(\ref{eq2}); the dotted line the factor
defined by formula (\ref{eq35}); the dash-dotted the factor
(\ref{eq33}). This figure demonstrates that the $S$-factors
considered have an essentially different behavior as $u\to1$.

To compare the relativistic factors~(\ref{eq32})
and~(\ref{eq35}) in  more detail, we plot in Fig.~\ref{S_fig3} the
ratio, which is denoted as ${N(\eta)}$, of the relativistic
factor~(\ref{eq32}) or~(\ref{eq35}) to the nonrelativistic
$S$-factor~(\ref{eq36}) for different values of $\alpha$ (the numbers
at the curves). The solid lines correspond to the
$S$-factor~(\ref{eq32}). The dashed lines, which are taken from
Ref.~\cite{YoonWong:0005}, represent the ratio of the
factor~(\ref{eq35}) to the nonrelativistic $S$-factor. As can be seen
from Fig.~\ref{S_fig3}, there is an essential difference between the
relativistic factors~(\ref{eq32}) and (\ref{eq35}). For example, in
the nonrelativistic limit, when $\eta$ increases (${\rm\,v}\to0$),
the relativistic $S$-factor~(\ref{eq35}) reproduces the
nonrelativistic limit only as $\alpha\to 0$ (see
Ref.~\cite{YoonWong:0005} for additional details).

Thus, the above analysis demonstrates that the relativistic
$S$-factor~(\ref{eq32}), as would be expected, coincides in form with
the nonrelativistic $S$-factor~(\ref{eq2}). However, the relative
velocity of an effective relativistic particle~(\ref{eq31}) emerging
instead of the system of two particles, now plays role of the
parameter of velocity, but not the relativistic relative velocity of
interacting particles, $\rm\,v$.

Let us briefly discuss the application of the $S$-factor. As noted
above in the introduction, in the vicinity of the quark-antiquark
threshold one cannot truncate the perturbative series and the
resummation factor should be taken into account in its entirety.
Involving a summation of threshold singularities by using the
$S$-factor leads to the following modification of the dominant
contribution of the function $R(s)$ in the case of vector
current~\footnote{The corresponding expression without the $S$-factor
can be found in Ref.~\cite{Reinders}.}
\begin{equation}
\label{eq37}
R(s) \to \,R_V^{(0)}(s)=\left[1-\frac{(m_1-m_2)^2}{s}\right]^2
\left[\frac{u \,(3-u^2)}{2} \,+ \, \frac{(m_1-m_2)^2}{2s}
\,u^3\right] S(u\,,\alpha),
\end{equation}
where, according to Eq.~(\ref{eq24}),
$ s=[(m_1+m_2)^2-(m_1-m_2)^2\,u^2]/(1-u^2)\,.$
By using this formula, we study the influence of the $S$-factor
to the function $R_V^{(0)}$.

\begin{figure}[htb]
\centering\includegraphics[width=0.6\textwidth]{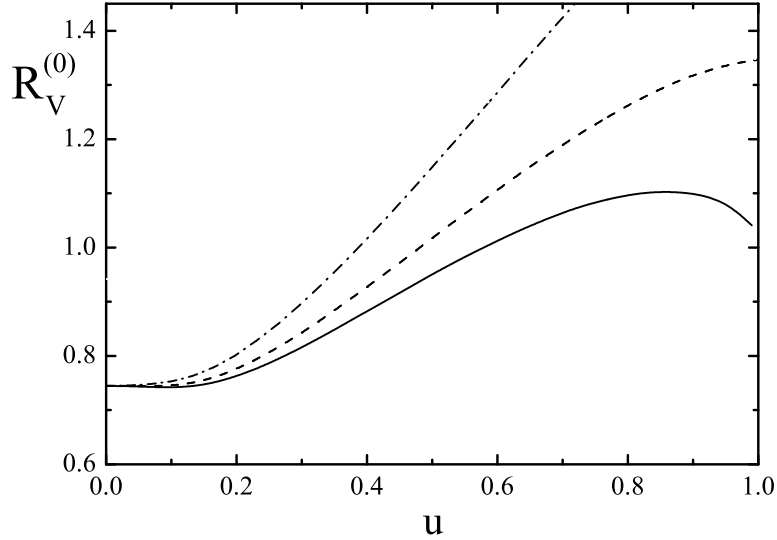}
 \caption{Behavior of the function $R_V^{(0)}$ given by
Eq.~(\ref{eq37}) as a function of variable $u$ for $\alpha=0.16$. The
result for the $S$-factor~(\ref{eq32}) is shown as the solid line,
for the nonrelativistic $S$-factor~(\ref{eq2}) as the dashed line,
and for the factor (\ref{eq33}) as the dash-dotted line.
 } \label{S_fig4}
\end{figure}

Figure~\ref{S_fig4} demonstrates the difference in the behavior of
$R_V^{(0)}$, which appears when we use the above factors, as
functions of variable $u$ for a fixed value of $\alpha=0.16$. The
solid line corresponds to the $S$-factor~(\ref{eq32}), the dashed
line to the nonrelativistic $S$-factor~(\ref{eq2}), and the
dashed-dotted line to expression~(\ref{eq33}). The curves are built
for effective quark masses $m_1=250$~MeV and $m_2=500$~MeV. These
values are close to the constituent masses $u$- and $s$-quarks (see,
for details, Ref.~\cite{MSS:06}).\footnote{We
plan to apply this $S$-factor to the hadronic $\tau$ decay mediated
by the strange current via $W^-\to s{\bar{u}}$.}
As can be seen from this figure,
the behavior of curves is essentially different, especially as
$u\to1$.

\begin{figure}[htb]
\centering\includegraphics[width=0.6\textwidth]{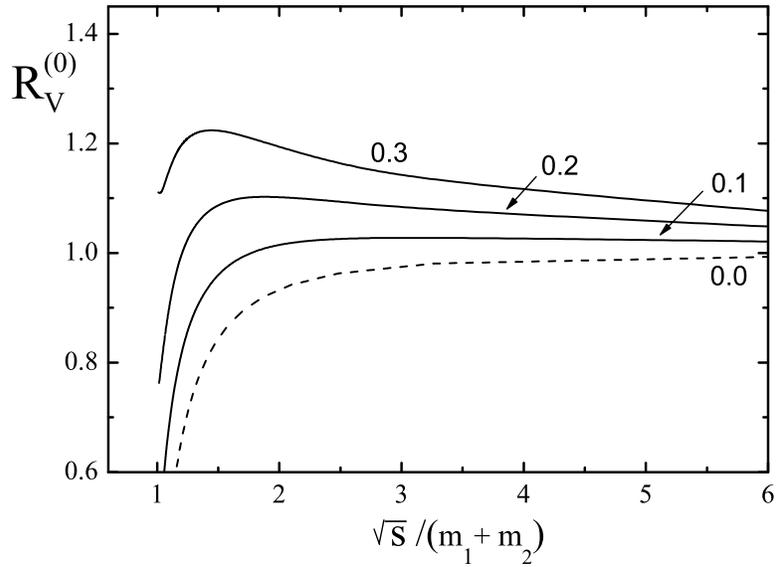}
\caption{Behavior of the function  $R_V^{(0)}$ with the
$S$-factor~(\ref{eq32}) as a function of dimensionless variable
$\sqrt{s}/(m_1+m_2)$ for different values of $\alpha$ (the numbers at
the curves). The dashed line represents $R_V^{(0)}$ without the
$S$-factor.} \label{S_fig5}
\end{figure}

Figure~\ref{S_fig5} demonstrates the behavior of the $R_V^{(0)}$
calculated with the new $S$-factor~(\ref{eq32}) obtained here versus
the dimensionless variable $\sqrt{s}/(m_1+m_2)$ for various values of
$\alpha$ (the numbers at the curves). The dashed line corresponds to
the case without the $S$-factor (or $\alpha=0$). This figure shows
that the influence of relativistic threshold resummation is much
stronger in the threshold region and with growing energy $\sqrt{s}$
weakens, and all curves approach unity.

\section{Conclusions}
In this paper, the new relativistic threshold resummation
$S$-factor~(\ref{eq32}) for the interaction of two relativistic
particles of unequal masses was obtained. For this aim the
relativistic quasipotential equation in relativistic configuration
representation~\cite{KadMM:70} with the Coulomb potential for the
interaction of two relativistic particles of unequal masses was used.
The Coulomb potential only formally has the same form as the
nonrelativistic potential but differs in the relativistic
configuration representation since its behavior corresponds to the
quark-antiquark potential $V_{q\bar{q}}\sim \bar{\alpha}_s (Q^2)/Q^2$
with the invariant charge $\bar{\alpha}_s(Q^2)\sim1/\ln{Q^2}$. So,
the principal effect coming from the running of the QCD coupling is
accumulated.

The new relativistic $S$-factor~(\ref{eq32}) obtained reproduces both
the known nonrelativistic behavior and the expected ultrarelativistic
limit. The new $S$-factor coincides in form with the nonrelativistic
$S$-factor~(\ref{eq2}); however, the role of the parameter of
velocity is played not by the relative velocity of interacting
particles, $\rm v$, but by the relative velocity~(\ref{eq31}) of an
effective relativistic particle emerging instead of the system of two
particles. It was shown that there is a difference (see
Figs.~\ref{S_fig2}--~\ref{S_fig4}) between the expression
(\ref{eq32}) obtained here and other known forms (\ref{eq32}),
(\ref{eq33}) and (\ref{eq34}). As the new relativistic resummation
factor (\ref{eq32}) was obtained within the framework of completely
covariant method, one can expect that this factor takes into account
more adequately relativistic nature of interaction.

It was demonstrated that the new relativistic resummation factor has
the influence on the behavior of the function $R(s)$. In some
physically interesting cases the function $R(s)$ occurs as a factor
in an integrand, as, for example, in the case of inclusive $\tau$
decay or in the Adler $D$-function, and the behavior of the
$S$-factor at intermediate values of variable $u$ becomes important.

\section*{Acknowledgments}
The authors would like to thank E.~A.~Kuraev, R.~N.~Faustov and
N.~B.~Skachkov for interest in this work and valuable discussions.

This work was supported in part by BelRFBR-JINR grant No.~F08D-001
and RFBR grant No.~08-01-00686.


\newpage

\centerline{FIGURES}

\vspace{5mm}

\begin{figure}[htb]
\centering\includegraphics[width=0.6\textwidth]{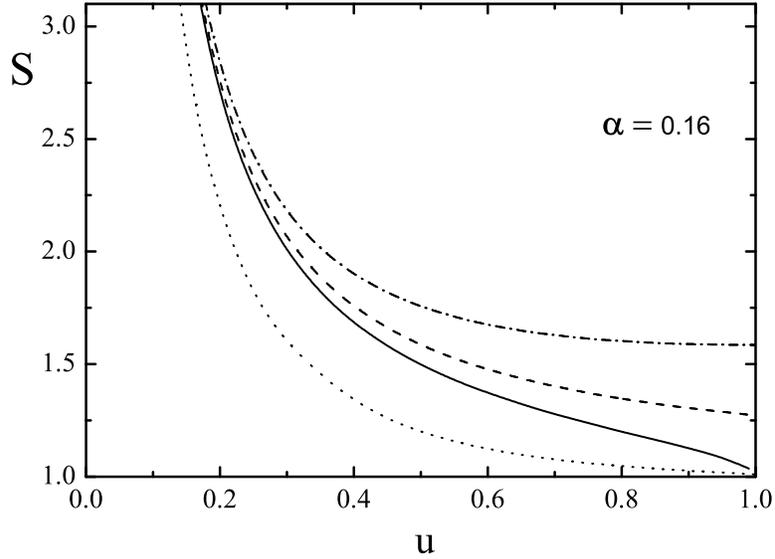}
\caption{Comparison of the $S$-factor behavior.
The solid curve corresponds
to the relativistic $S$-factor~(\ref{eq32}). The nonrelativistic
$S$-factor~(\ref{eq2}) is presented by the dashed line, the factor
(\ref{eq33}) is presented by the dash-dotted line, and the factor
(\ref{eq35}) is shown as the dotted line.}
\label{S_fig2}
\end{figure}

\begin{figure}[htb]
\centering\includegraphics[width=0.65\textwidth]{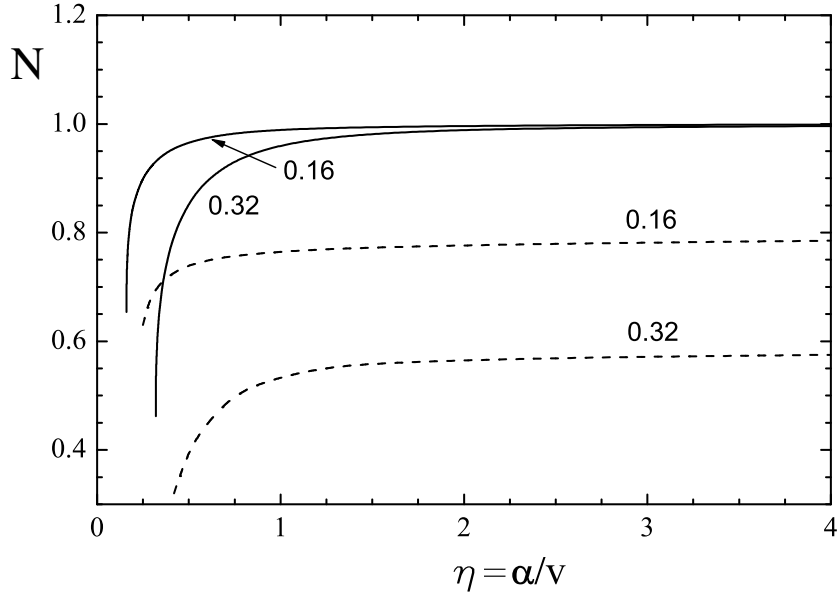}
\caption{Ratio of relativistic $S$-factor to nonrelativistic one,
$N(\eta)$, for different values of $\alpha$ (the numbers at the
curves). The solid lines correspond to the $S$-factor~(\ref{eq32})
and the dashed lines taken from Ref.~\cite{YoonWong:0005} correspond
to the factor~(\ref{eq35}).} \label{S_fig3}
\end{figure}

\end{document}